\documentclass[utf8]{frontiersFPHY} 

\setcitestyle{square,authoryear} 
\usepackage{url,hyperref,lineno,microtype,subcaption}
\usepackage[onehalfspacing]{setspace}

\def\keyFont{\fontsize{8}{11}\helveticabold }
\def\firstAuthorLast{Venturi {et~al.}} 
\def\Authors{G. Venturi\,$^{1,2,*}$, A. Marconi\,$^{1,2}$, M. Mingozzi\,$^{2,3}$, S. Carniani\,$^{4,5}$, G. Cresci\,$^{2}$, G. Risaliti\,$^{1,2}$ and F. Mannucci\,$^{2}$}
\def\Address{$^{1}$Dipartimento di Fisica e Astronomia, Università degli Studi di Firenze, Sesto Fiorentino (Firenze), Italy \\
$^{2}$Osservatorio Astrofisico di Arcetri, INAF, Firenze, Italy \\
$^{3}$Dipartimento di Fisica e Astronomia, Università di Bologna, Bologna, Italy \\
$^{4}$Cavendish Laboratory, Department of Physics, University of Cambridge, Cambridge, UK \\
$^{5}$Kavli Institute for Cosmology, University of Cambridge, Cambridge, UK}
\def\corrAuthor{Giacomo Venturi}

\def\corrEmail{Osservatorio Astrofisico di Arcetri, INAF, Largo E. Fermi 5, Firenze, I-50125, Italy, gventuri@arcetri.astro.it}

\begin{document}
\onecolumn
\firstpage{1}

\title[Ionized gas outflows from the MAGNUM survey]{Ionized gas outflows from the MAGNUM survey: NGC 1365 and NGC 4945} 

\author[\firstAuthorLast ]{\Authors} 
\address{} 
\correspondance{} 

\extraAuth{}

\maketitle

\begin{abstract}

\section{}
AGN feedback, acting through strong outflows accelerated in the nuclear region of AGN hosts, is invoked as a key ingredient for galaxy evolution by many models to explain the observed BH-galaxy scaling relations. Recently, some direct observational evidence of radiative mode feedback in action has been finally found in quasars at $z>$1.5. However, it is not possible to study outflows in quasars at those redshifts on small scales ($\lesssim$100 pc), as spatial information is limited by angular resolution. This is instead feasible in nearby active galaxies, which are ideal laboratories to explore outflow structure and properties, as well as the effects of AGN on their host galaxies. In this proceeding we present preliminary results from the MAGNUM survey, which comprises nearby Seyfert galaxies observed with the integral field spectrograph VLT/MUSE. We focus on two sources, NGC 1365 and NGC 4945, that exhibit double conical outflows extending on distances $>$1 kpc. We disentangle the dominant contributions to ionization of the various gas components observed in the central $\sim$5.3 kpc of NGC 1365. An attempt to infer outflow 3D structure in NGC 4945 is made via simple kinematic modeling, suggesting a hollow cone geometry.

%

\tiny
 \keyFont{ \section{Keywords:} active galactic nuclei, galaxies, outflows, NGC 1365, NGC 4945, imaging spectroscopy} 
\end{abstract}

\section{Introduction}
AGN are believed to have a strong influence on their host galaxies, accelerating fast outflows able to quench star formation (negative feedback). Many models invoke AGN feedback as a key ingredient for galaxy evolution (e.g. \citealt{Springel:2005aa}, \citealt{Hopkins:2006aa}, \citealt{Ciotti:2010aa}, \citealt{Scannapieco:2012aa}), shaping galaxy properties and giving rise to the observed black hole-host galaxy relations ($M_{\textrm{BH}}$-$\sigma_{\textrm{bulge}}$, $M_{\textrm{BH}}$-$L_{\textrm{bulge}}$, $M_{\textrm{BH}}$-$M_{\textrm{bulge}}$; e.g. \citealt{Gebhardt:2000aa}, \citealt{Ferrarese:2000aa}, \citealt{Marconi:2003aa}, \citealt{McConnell:2013aa}, \citealt{Kormendy:2013aa}). According to radiative (or wind) mode feedback models, AGN at the Eddington limit can sweep away surrounding gas by radiation pressure, driving fast outflows which overcome the gravitational potential of the galaxy. Momentum balance in radiative mode gives a $M_{\textrm{BH}}$-$\sigma_{\textrm{bulge}}$ relation which is in agreement with the observed one (e.g. \citealt{Fabian:1999aa}, \citealt{King:2003aa, King:2005aa}, \citealt{Murray:2005aa}).

Luminous AGN are the best candidates to detect ongoing star formation being quenched by AGN feedback, as indeed found in recent works (e.g. \citealt{Cano-Diaz:2012aa}, \citealt{Cresci:2015aa}, \citealt{Brusa:2015aa}, \citealt{Carniani:2016aa}, \citealt{Carniani:2017aa}). However, it is not possible to study high-$z$ quasar outflows on small spatial scales ($\lesssim$100 pc), due to their large distances (e.g. even with adaptive optics \citealt{Williams:2017aa} achieve a spatial resolution of $\sim$1 kpc at z=2.4). On the contrary, nearby active galaxies are ideal laboratories to explore in detail outflow properties, their formation and acceleration mechanisms, as well as the effects of AGN activity on host galaxies (e.g. \citealt{Garcia-Burillo:2014aa}, \citealt{Storchi-Bergmann:2015aa}, \citealt{Cresci:2015ab}, \citealt{Davies:2016aa}). Here we present preliminary results from our MAGNUM survey (Measuring Active Galactic Nuclei Under MUSE Microscope), which aims at investigating the properties of outflows in nearby AGN, the physical conditions of the ionized gas and the interplay between nuclear activity and star formation in the host galaxy, thanks to the unprecedented combination of spatial and spectral coverage provided by the integral field spectrograph MUSE at VLT (1$'\times$1$'$ FOV with 0.2$''$ per spaxel and 4750-9300 \AA\ spectral coverage; \citealt{Bacon:2010aa}). The first result from the MAGNUM survey is presented in \citealt{Cresci:2015ab}, where tentative evidence of star formation induced by the AGN outflow (positive feedback; e.g. \citealt{Silk:2013aa}) is found in the nearby Seyfert galaxy NGC 5643. Here two young ($\sim$10$^7$ yr) isolated star-forming clumps are observed in the direction of the AGN-ionized outflow, where this impacts the dense material at the edge of the dust lane, suggesting a case of occurring positive feedback.

\section{MAGNUM survey: overview}
\subsection{Sample selection}
MAGNUM galaxies have been picked out from the optically-selected samples of \citealt{Maiolino:1995aa} and \citealt{Risaliti:1999aa} (MR95 and R99, respectively, from now on) and from Swift-BAT 70-month Hard X-ray Survey (\citealt{Baumgartner:2013aa}; hereafter SB-70m). The latter includes optically-obscured AGN which are excluded by the constraints given in MR95 and R99. On the other hand, a hard-X ray selection, as it is SB-70m, misses Compton-thick AGN, embraced instead by a selection based on optical emission lines (like MR95 and R99), which are spatially extended and not confined to the nuclear spot as the hard X-ray emission. We chose only those sources which are observable from Paranal Observatory ($-70^{\circ} < \delta < 20^{\circ}$) and have a distance $<$50 Mpc, ending up with a total of 73 objects. 

So far, we have analyzed MUSE data of 10 galaxies belonging to the MAGNUM sample,  comprising Centaurus A, Circinus, IC 5063, NGC 1068, NGC 1365, NGC 1386, NGC 2992, NGC 4945, NGC 5643 and NGC 6810.

\subsection{Data analysis}
We have fitted and subtracted the stellar continuum, after having Voronoi-binned (\citealt{Cappellari:2003aa}) the data cube so as to get an average signal-to-noise ratio per wavelength channel (1.25 \AA/channel) on the stellar continuum of at least 50 in each bin. We made use of a linear combination of \citealt{Vazdekis:2010aa} synthetic spectral energy distributions (SEDs) for single-age, single-metallicity stellar populations (SSPs), employing pPXF code (Penalized Pixel-Fitting; \citealt{Cappellari:2004aa}) to convolve the linearly combined stellar templates with a Gaussian profile so as to reproduce the systemic velocity and the velocity dispersion of the stellar absorption lines. The resulting star-subtracted cube has then been spatially smoothed with a 1$\sigma$-px Gaussian kernel. All the main emission lines have been fitted, spaxel-by-spaxel, with 1, 2 or 3 Gaussian components to best reproduce the total line profiles. The number of components in the best fits is set by the value of the reduced $\chi^2$ with the aim of providing the best fit with the minimum number of free parameters. The same number of components is used for all emission lines imposing the same velocity and width to each component.

All the details about the MAGNUM sample selection and the data analysis will be given in Venturi et al. 2018, in prep.

\section{MUSE maps}
Ionized outflows are observed in all the galaxies of our sample, with a more or less defined conical shape and a complex kinematics. In the following, we present the representative cases of NGC 1365 and NGC 4945. The former was observed with MUSE on October 12, 2014, for a total exposure time of $\sim$4000 s, the latter on January 17, 2015, for a total of $\sim$2000 s. NGC 1365 was studied in the ionized gas in many articles, e.g.\ \citealt{Storchi-Bergmann:1991aa}, \citealt{Veilleux:2003aa}, \citealt{Sharp:2010aa}. \citealt{Lindblad:1999aa} gives a comprehensive review about the early work on this galaxy. An analysis of the ionized gas in NGC 4945 is reported e.g.\ in \citealt{Marconi:2000aa}, \citealt{Rossa:2003aa}.

\subsection{NGC 1365}
\begin{figure}[b]
\begin{center}
\includegraphics[trim={0 3cm 0 3cm},clip,width=\textwidth]{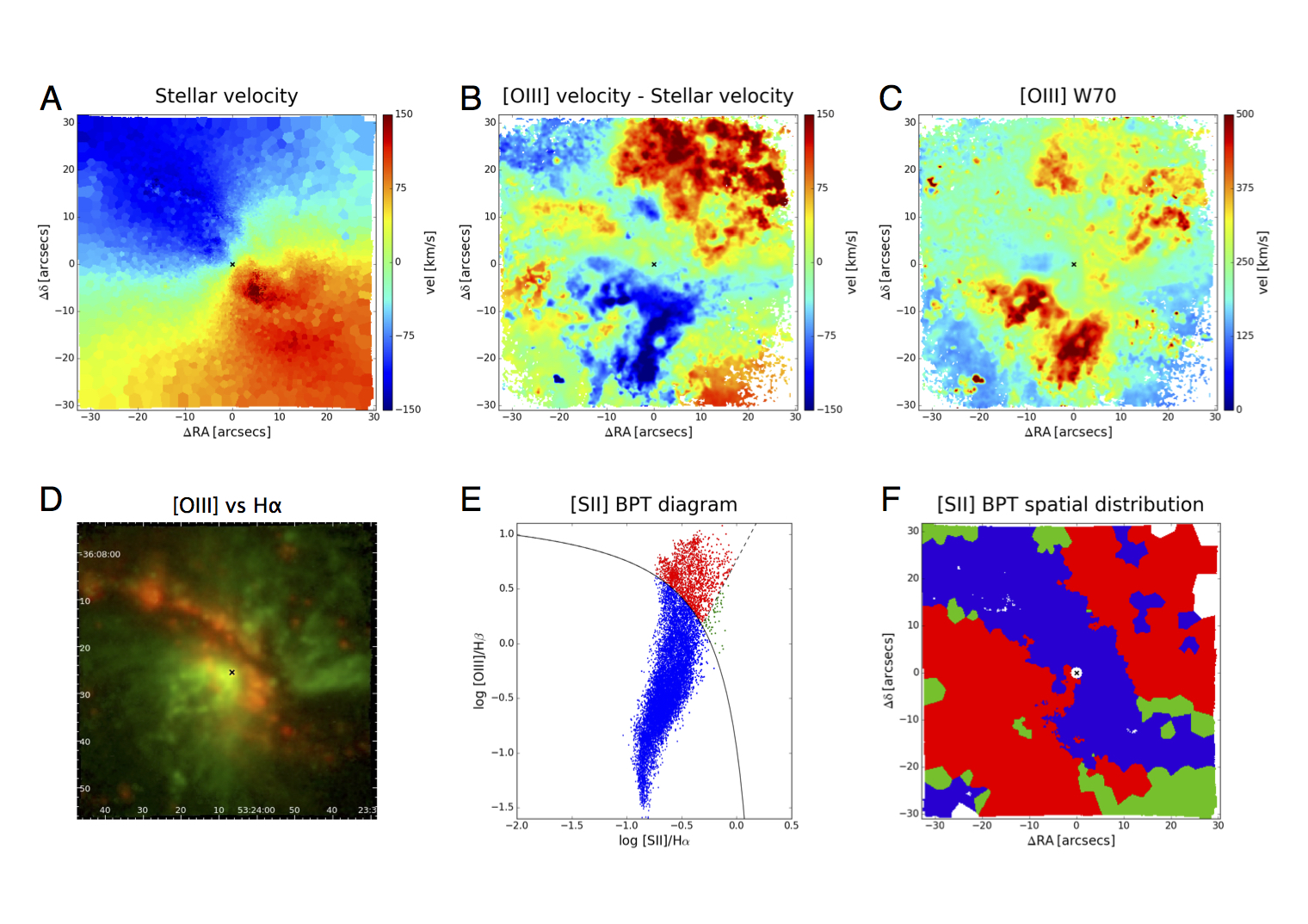}
\end{center}
\caption{Maps from our analysis of MUSE observations of NGC 1365. North is up. The cross marks the active nucleus. [OIII] kinematic maps in (B) and (C) are 1px-$\sigma$ spatially re-smoothed to get a better visual output. (A) Stellar velocity map, w.r.t.\ the systemic velocity of 1630 km/s adopted. The map has been obtained from the spectral shift of the stellar absorption lines resulting from the fit of the stellar emission carried out on the Voronoi-binned data cube, for which an average signal-to-noise ratio per wavelength channel of 50 has been requested in each bin. (B) Difference between the [OIII] velocity and the stellar velocity, in order to isolate the proper motions of the gas w.r.t.\ the stars. The [OIII] velocity is the first-order moment of the total fitted line profile. (C) [OIII] W70 map, i.e. the difference between the velocities at the 85th and at the 15th percentile of the total fitted line profile. (D) Two-color image of [OIII] (green) and H$\alpha$ (red) integrated emission of the total fitted line profile. (E) [SII] $\lambda\lambda$6716,6731/H$\alpha$ vs. [OIII] $\lambda$5007/H$\beta$ spatially resolved BPT diagram, obtained from the fit of the star-subtracted Voronoi-binned cube (so as to have an average signal-to-noise ratio per wavelength channel around H$\beta$ of at least 4 in each bin). Each point in the diagram then corresponds to a bin in the FOV. The solid curve marks \citealt{Kewley:2001aa} theoretical upper bound for pure star formation, while the dashed line separates Seyferts from LI(N)ERs (\citealt{Kewley:2006aa}). Blue points are then star-formation dominated bins, red ones are AGN-dominated ones, while green one have LI(N)ER-like ratios. (F) Spatial distribution of the bins, color-coded according to the spatially resolved BPT diagram in (E).}\label{fig:1}
\end{figure}

NGC 1365 is a barred spiral Seyfert galaxy at $z$ = 0.00546, extending for 11.2$'\times$6.2$'$, so that the $\sim$1$'\times$1$'$ FOV of MUSE covers the central $\sim$5.3$\times$5.3 kpc$^2$ of the galaxy.
In figures \ref{fig:1}A, \ref{fig:1}B and \ref{fig:1}C we show the stellar velocity field, the average velocity of the [OIII] $\lambda 5007$ line ([OIII] hereafter) subtracted, spaxel-by-spaxel, by the stellar velocity and the map of [OIII] line width W70 (i.e.\ the difference between the velocities at the 85th and at the 15th percentile of the total fitted profile; see, e.g., \citealt{Harrison:2014aa}), respectively. The first one exhibits the rotational motion of the stars (approaching to the NE, receding to the SW), which has a twisted shape probably associated with the presence of the bar. We subtract the average [OIII] velocity by the stellar one so as to isolate gas motions deviating from stellar rotation. The two kinematic maps of [OIII] reveal a double conical outflow with a clumpy morphology, oriented in the direction SE-NW, where [OIII] velocity deviates most from stellar one (by over 150 km/s) and total line profile is broader compared to the rest of the FOV, giving rise to W70 values greater than 500 km/s.
The broadening is due to the fact that in each spaxel all the kinematic components in the line of sight, the one from the outflow and the one from the disk, sum up giving a complex line profile. Double-peaked line profiles are indeed ubiquitous in the regions having higher W70. As shown by the velocity map the SE cone is approaching to the observer, while the NW one is receding.

The outflow is part of the [OIII]-emitting cone, which can be seen in figure \ref{fig:1}D in green, where H$\alpha$, tracing star formation in the disk, is instead reported in red. Both represent the integrated flux of the total fitted line profile. The spatial distribution of the two emission lines is clearly different, with the H$\alpha$ being dominant in the direction NE-SW (along the galactic bar), almost perpendicular to the [OIII] double cone. BPT diagnostic diagrams (\citealt{Baldwin:1981aa}, \citealt{Veilleux:1987aa}) give a quantitative assessment regarding the dominant ionizing source.
In figure \ref{fig:1}E we report the spatially resolved [SII] $\lambda\lambda$6716,6731/H$\alpha$ BPT diagram of NGC 1365, generated, differently from the other maps, from the fit of another data cube, produced by performing a Voronoi binning on the star-subtracted one, so as to get an average signal-to-noise ratio per wavelength channel around H$\beta$ of at least 4 in each bin. This allows to get more reliable line ratios for the diagnostic diagram.
Each point in the BPT corresponds to a single bin in the FOV and they are color-coded in the following way: blue points mark star-forming dominated bins, standing below \citealt{Kewley:2001aa} upper limit for pure star formation (solid curve), while red ones indicate AGN-dominated bins, separated by \citealt{Kewley:2006aa} dashed line from the green ones having LI(N)ER-like line ratios. The spatial distribution of the bins in the FOV is presented in figure \ref{fig:1}F, using the same color coding adopted in the diagram. The diagonal lane in the direction NE-SW is dominated by ionization from OB stars in star-forming regions, while the almost perpendicular [OIII]-emitting cone is AGN-dominated, indicating that the ionized outflow is AGN-driven. Besides the diagonal lane, some isolated bins far from the center, emerging from the cone, have star-forming ratios as well and correspond to the strong red H$\alpha$ blobs standing out against the green [OIII] in figure \ref{fig:1}C. A circular area surrounding the nucleus has been masked due to the presence of BLR lines disturbing the fit and giving highly deviant values of the line ratios.

The maps in figure \ref{fig:1} then suggest that the SE lobe of the AGN-ionized cones is the nearest to the observer and stands above the disk while the NW one is the furthest and resides behind the disk, as the former shows approaching velocities and the strongest flux between the two, especially going closer to the center, while the latter recedes and is fainter and overcome by the dusty star-forming regions near the center.

\subsection{NGC 4945}

\begin{figure}[b]
\begin{center}
\includegraphics[trim={0 9cm 0 0},clip,width=\textwidth]{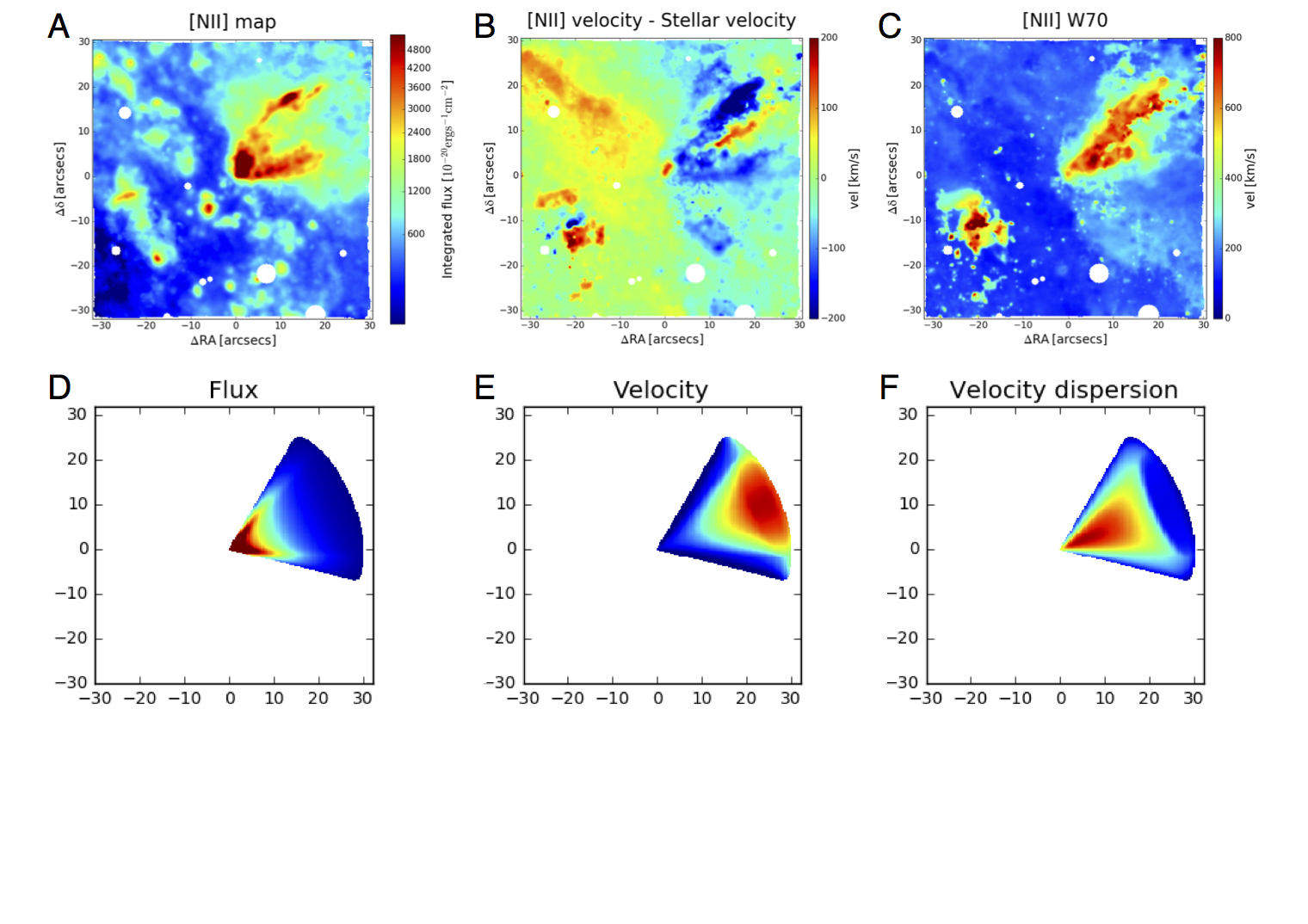}
\end{center}
\caption{Maps from our analysis of MUSE data of NGC 4945 (top) and from kinematic model (bottom). North is up. [NII] maps on top are 1px-$\sigma$ spatially re-smoothed, so as to get a better visual output. White circles indicate masked regions coinciding with foreground stars. (A) [NII] flux map. (B) [NII] velocity spaxel-by-spaxel subtracted by the stellar velocity, in order to isolate proper motions of the gas w.r.t.\ stars. The [NII] velocity is the first-order moment of the total fitted line profile. (C) [NII] W70 map, i.e difference between the 85th-percentile velocity and the 15th-percentile one calculated on the fitted line profile. (D) Flux map from simple model of hollow cone having inclination of 75$^{\circ}$ w.r.t. line of sight, inner and outer half opening angle of 25$^{\circ}$ and 35$^{\circ}$, respectively (thus intercepting the plane of the sky), and constant velocity field. (E) Velocity map from model in (D). (F) Velocity dispersion map from model in (D). Data maps in (A), (B), (C) and model maps in (D), (E), (F) have matching color scales.}\label{fig:2}
\end{figure}

A double conical outflow is observed in another source belonging to our MAGNUM survey, i.e. NGC 4945, an almost perfectly edge-on galaxy at $z$ = 0.00188. [NII] total integrated flux, average velocity (spaxel-by-spaxel subtracted by the stellar velocity) and W70 maps of this galaxy obtained from the star-subtracted spatially-smoothed data cube are presented in figures \ref{fig:2}A, \ref{fig:2}B and \ref{fig:2}C, respectively. The side of the FOV corresponds to $\sim$3.5 kpc. The presence of the outflow is clear in W70 map, where a NW lobe and a SE one can be observed, the former being more extended than the latter, reaching a distance of $\sim$1.8 kpc from the center. In the flux map too the NW lobe is much stronger and more extended than the SE one. The galaxy is nearly edge on and dusty, in fact, and so it is plausible that the galactic plane resides behind the NW lobe w.r.t. the observer, while the SE one stands behind the disk and is obscured by dust close to the center but emerges from it at a certain point. The velocity map of [NII] shows the kinematic structure of the two cones, the NW one having approaching velocities at its edges and receding ones around its axis, the SE lobe having an opposite behavior, i.e. receding velocities at its edges and approaching ones around its axis. As in the case of NGC 1365, double-peaked and complicated line profiles are ubiquitous along the outflow.

The aspect of the velocity map of the NW cone could be explained thinking to a hollow cone which has an axis inclination w.r.t. the l.o.s. $<$90$^{\circ}$ and a sufficient aperture such that the far part of the cone intercepts the plane of the sky (the same holds for the SE lobe, but with opposite geometry). Another feature supporting this scenario is the flux enhancement at the edges of [NII] cone in figure \ref{fig:2}A, compatible with an effect of limb brightening in a hollow cone. To test this hypothesis a simple kinematic model has been realized. The model features a hollow cone having inclination w.r.t. line of sight of 75$^{\circ}$, inner and outer half opening angle of 25$^{\circ}$ and 35$^{\circ}$, respectively, and constant velocity field. The resulting maps of flux, velocity and velocity dispersion are reported in figures \ref{fig:2}D, \ref{fig:2}E and \ref{fig:2}F, respectively. The velocity map of the model resembles the observed one, with approaching velocities at the edges of the cone and receding ones in the middle, and also the velocity dispersion spatial distribution seems to reproduce well the observed one. Although this simple model does not take into account the clumpiness of the ionized gas, it seems to be a fairly good representation of the observed maps, and thus a hollow cone intercepting the plane of the sky appears to be a promising way to explain the complex kinematics that we observe in NGC 4945. We are developing a more detailed and sophisticated kinematic model to actually reconstruct outflow 3D shape and intrinsic properties.

\section{Conclusions}
We presented a study of ionized outflows in nearby AGN in the field of our MAGNUM survey, which takes advantage of the unprecedented capabilities of MUSE at VLT to study in detail outflows, photoionization and the interplay between AGN and host galaxies. We first gave a brief overview of the MAGNUM survey, describing the sample selection and the standard data analysis carried out on the 10 targets we studied up to now. We then presented preliminary maps and results obtained for two targets of the MAGNUM survey, NGC 1365 and NGC 4945.

MUSE observations of NGC 1365 reveal a clumpy double conical outflow, propagating from the center in the SE-NW direction, the brighter SE lobe approaching, the fainter NW one receding. The outflow is almost perpendicular to the stellar rotational field and is part of the [OIII]-emitting double cone, whose ionization is dominated by the AGN continuum, as established by [SII] $\lambda\lambda$6716,6731/H$\alpha$ vs. [OIII] $\lambda$5007/H$\beta$ spatially resolved BPT diagram. The BPT also shows that the diagonal elongated area of the galaxy following the bar direction, nearly perpendicular to the AGN-ionized double cone, is instead dominated by ionization in star-forming regions. The fact that the SE cone is approaching and is brighter w.r.t. the NW receding one, in particular near the center where the star formation is dominant, suggests that the NW cone is partially hidden behind the galaxy disk, especially near the central dusty region, while the SE one points towards the observer above the disk.

NGC 4945 shows a double conical outflow as well, the NW lobe being much brighter than the SE one, which appears only at $\sim$15$''$ from the center, likely being completely dust-obscured at smaller radius, as the galaxy is almost edge-on. The kinematic structure of the NW cone reveals approaching velocities at the edges of the cone and receding ones along its axis. A simple kinematic model of a hollow cone pointing towards the observer's side but intercepting the plane of the sky accounts for this observed field. Consistently, the SE lobe shows an opposite behavior, with receding velocities at the edges of the cone and approaching ones along its axis.

\section*{Conflict of Interest Statement}

The authors declare that the research was conducted in the absence of any commercial or financial relationships that could be construed as a potential conflict of interest.

\section*{Author Contributions}

GV wrote the article, GV performed the data reduction using both the ESO pipeline and python scripts written by SC, AM wrote the first version of the python scripts for data analysis, GV and MM tested and implemented the scripts and performed the analysis of the sources, finally producing the maps, all the co-authors provided a critical review to the article.


\section*{Funding}
SC acknowledges financial support from the Science and Technology Facilities Council (STFC).



\bibliographystyle{frontiersinHLTH&FPHY} 
\bibliography{bibliography_v1}

\begin{thebibliography}{42}
\expandafter\ifx\csname natexlab\endcsname\relax\def\natexlab#1{#1}\fi
\expandafter\ifx\csname urlstyle\endcsname\relax
  \expandafter\ifx\csname doi\endcsname\relax
  \def\doi#1{doi:\discretionary{}{}{}#1}\fi \else
  \expandafter\ifx\csname doi\endcsname\relax
  \def\doi{doi:\discretionary{}{}{}\begingroup \urlstyle{rm}\Url}\fi \fi
\expandafter\ifx\csname selectlanguage\endcsname\relax
  \def\selectlanguage#1{}\fi

\bibitem[{{Springel} et~al.(2005){Springel}, {Di Matteo}, and
  {Hernquist}}]{Springel:2005aa}
{Springel} V, {Di Matteo} T, {Hernquist} L.
\newblock {Modelling feedback from stars and black holes in galaxy mergers}.
\newblock {\em \mnras\/} {\bf 361} (2005) 776--794.
\newblock \doi{10.1111/j.1365-2966.2005.09238.x}.

\bibitem[{{Hopkins} et~al.(2006){Hopkins}, {Hernquist}, {Cox}, {Di Matteo},
  {Robertson}, and {Springel}}]{Hopkins:2006aa}
{Hopkins} PF, {Hernquist} L, {Cox} TJ, {Di Matteo} T, {Robertson} B, {Springel}
  V.
\newblock {A Unified, Merger-driven Model of the Origin of Starbursts, Quasars,
  the Cosmic X-Ray Background, Supermassive Black Holes, and Galaxy Spheroids}.
\newblock {\em \apjs\/} {\bf 163} (2006) 1--49.
\newblock \doi{10.1086/499298}.

\bibitem[{{Ciotti} et~al.(2010){Ciotti}, {Ostriker}, and
  {Proga}}]{Ciotti:2010aa}
{Ciotti} L, {Ostriker} JP, {Proga} D.
\newblock {Feedback from Central Black Holes in Elliptical Galaxies. III.
  Models with Both Radiative and Mechanical Feedback}.
\newblock {\em \apj\/} {\bf 717} (2010) 708--723.
\newblock \doi{10.1088/0004-637X/717/2/708}.

\bibitem[{{Scannapieco} et~al.(2012){Scannapieco}, {Wadepuhl}, {Parry},
  {Navarro}, {Jenkins}, {Springel} et~al.}]{Scannapieco:2012aa}
{Scannapieco} C, {Wadepuhl} M, {Parry} OH, {Navarro} JF, {Jenkins} A,
  {Springel} V, et~al.
\newblock {The Aquila comparison project: the effects of feedback and numerical
  methods on simulations of galaxy formation}.
\newblock {\em \mnras\/} {\bf 423} (2012) 1726--1749.
\newblock \doi{10.1111/j.1365-2966.2012.20993.x}.

\bibitem[{{Gebhardt} et~al.(2000){Gebhardt}, {Bender}, {Bower}, {Dressler},
  {Faber}, {Filippenko} et~al.}]{Gebhardt:2000aa}
{Gebhardt} K, {Bender} R, {Bower} G, {Dressler} A, {Faber} SM, {Filippenko} AV,
  et~al.
\newblock {A Relationship between Nuclear Black Hole Mass and Galaxy Velocity
  Dispersion}.
\newblock {\em \apjl\/} {\bf 539} (2000) L13--L16.
\newblock \doi{10.1086/312840}.

\bibitem[{{Ferrarese} and {Merritt}(2000)}]{Ferrarese:2000aa}
{Ferrarese} L, {Merritt} D.
\newblock {A Fundamental Relation between Supermassive Black Holes and Their
  Host Galaxies}.
\newblock {\em \apjl\/} {\bf 539} (2000) L9--L12.
\newblock \doi{10.1086/312838}.

\bibitem[{{Marconi} and {Hunt}(2003)}]{Marconi:2003aa}
{Marconi} A, {Hunt} LK.
\newblock {The Relation between Black Hole Mass, Bulge Mass, and Near-Infrared
  Luminosity}.
\newblock {\em \apjl\/} {\bf 589} (2003) L21--L24.
\newblock \doi{10.1086/375804}.

\bibitem[{{McConnell} and {Ma}(2013)}]{McConnell:2013aa}
{McConnell} NJ, {Ma} CP.
\newblock {Revisiting the Scaling Relations of Black Hole Masses and Host
  Galaxy Properties}.
\newblock {\em \apj\/} {\bf 764} (2013) 184.
\newblock \doi{10.1088/0004-637X/764/2/184}.

\bibitem[{{Kormendy} and {Ho}(2013)}]{Kormendy:2013aa}
{Kormendy} J, {Ho} LC.
\newblock {Coevolution (Or Not) of Supermassive Black Holes and Host Galaxies}.
\newblock {\em \araa\/} {\bf 51} (2013) 511--653.
\newblock \doi{10.1146/annurev-astro-082708-101811}.

\bibitem[{{Fabian}(1999)}]{Fabian:1999aa}
{Fabian} AC.
\newblock {The obscured growth of massive black holes}.
\newblock {\em \mnras\/} {\bf 308} (1999) L39--L43.
\newblock \doi{10.1046/j.1365-8711.1999.03017.x}.

\bibitem[{{King}(2003)}]{King:2003aa}
{King} A.
\newblock {Black Holes, Galaxy Formation, and the M$_{BH}$-{$\sigma$}
  Relation}.
\newblock {\em \apjl\/} {\bf 596} (2003) L27--L29.
\newblock \doi{10.1086/379143}.

\bibitem[{{King}(2005)}]{King:2005aa}
{King} A.
\newblock {The AGN-Starburst Connection, Galactic Superwinds, and
  M$_{BH}$-{$\sigma$}}.
\newblock {\em \apjl\/} {\bf 635} (2005) L121--L123.
\newblock \doi{10.1086/499430}.

\bibitem[{{Murray} et~al.(2005){Murray}, {Quataert}, and
  {Thompson}}]{Murray:2005aa}
{Murray} N, {Quataert} E, {Thompson} TA.
\newblock {On the Maximum Luminosity of Galaxies and Their Central Black Holes:
  Feedback from Momentum-driven Winds}.
\newblock {\em \apj\/} {\bf 618} (2005) 569--585.
\newblock \doi{10.1086/426067}.

\bibitem[{{Cano-D{\'{\i}}az} et~al.(2012){Cano-D{\'{\i}}az}, {Maiolino},
  {Marconi}, {Netzer}, {Shemmer}, and {Cresci}}]{Cano-Diaz:2012aa}
{Cano-D{\'{\i}}az} M, {Maiolino} R, {Marconi} A, {Netzer} H, {Shemmer} O,
  {Cresci} G.
\newblock {Observational evidence of quasar feedback quenching star formation
  at high redshift}.
\newblock {\em \aap\/} {\bf 537} (2012) L8.
\newblock \doi{10.1051/0004-6361/201118358}.

\bibitem[{{Cresci} et~al.(2015{\natexlab{a}}){Cresci}, {Mainieri}, {Brusa},
  {Marconi}, {Perna}, {Mannucci} et~al.}]{Cresci:2015aa}
{Cresci} G, {Mainieri} V, {Brusa} M, {Marconi} A, {Perna} M, {Mannucci} F,
  et~al.
\newblock {Blowin' in the Wind: Both ``Negative'' and ``Positive'' Feedback in
  an Obscured High-z Quasar}.
\newblock {\em \apj\/} {\bf 799} (2015{\natexlab{a}}) 82.
\newblock \doi{10.1088/0004-637X/799/1/82}.

\bibitem[{{Brusa} et~al.(2015){Brusa}, {Feruglio}, {Cresci}, {Mainieri},
  {Sargent}, {Perna} et~al.}]{Brusa:2015aa}
{Brusa} M, {Feruglio} C, {Cresci} G, {Mainieri} V, {Sargent} MT, {Perna} M,
  et~al.
\newblock {Evidence for feedback in action from the molecular gas content in
  the z \~{} 1.6 outflowing QSO XID2028}.
\newblock {\em \aap\/} {\bf 578} (2015) A11.
\newblock \doi{10.1051/0004-6361/201425491}.

\bibitem[{{Carniani} et~al.(2016){Carniani}, {Marconi}, {Maiolino},
  {Balmaverde}, {Brusa}, {Cano-D{\'{\i}}az} et~al.}]{Carniani:2016aa}
{Carniani} S, {Marconi} A, {Maiolino} R, {Balmaverde} B, {Brusa} M,
  {Cano-D{\'{\i}}az} M, et~al.
\newblock {Fast outflows and star formation quenching in quasar host galaxies}.
\newblock {\em \aap\/} {\bf 591} (2016) A28.
\newblock \doi{10.1051/0004-6361/201528037}.

\bibitem[{{Carniani} et~al.(2017){Carniani}, {Marconi}, {Maiolino}, {Feruglio},
  {Brusa}, {Cresci} et~al.}]{Carniani:2017aa}
{Carniani} S, {Marconi} A, {Maiolino} R, {Feruglio} C, {Brusa} M, {Cresci} G,
  et~al.
\newblock {AGN feedback on molecular gas reservoirs in quasars at $z\sim$2.4}.
\newblock {\em ArXiv e-prints\/}  (2017).

\bibitem[{{Williams} et~al.(2017){Williams}, {Maiolino}, {Krongold},
  {Carniani}, {Cresci}, {Mannucci} et~al.}]{Williams:2017aa}
{Williams} RJ, {Maiolino} R, {Krongold} Y, {Carniani} S, {Cresci} G, {Mannucci}
  F, et~al.
\newblock {An ultra-dense fast outflow in a quasar at z = 2.4}.
\newblock {\em \mnras\/} {\bf 467} (2017) 3399--3412.
\newblock \doi{10.1093/mnras/stx311}.

\bibitem[{{Garc{\'{\i}}a-Burillo} et~al.(2014){Garc{\'{\i}}a-Burillo},
  {Combes}, {Usero}, {Aalto}, {Krips}, {Viti} et~al.}]{Garcia-Burillo:2014aa}
{Garc{\'{\i}}a-Burillo} S, {Combes} F, {Usero} A, {Aalto} S, {Krips} M, {Viti}
  S, et~al.
\newblock {Molecular line emission in NGC 1068 imaged with ALMA. I. An
  AGN-driven outflow in the dense molecular gas}.
\newblock {\em \aap\/} {\bf 567} (2014) A125.
\newblock \doi{10.1051/0004-6361/201423843}.

\bibitem[{{Storchi Bergmann}(2015)}]{Storchi-Bergmann:2015aa}
{Storchi Bergmann} T.
\newblock {The Narrow Line Region in 3D: mapping AGN feeding and feedback}.
\newblock {Ziegler} BL, {Combes} F, {Dannerbauer} H, {Verdugo} M, editors, {\em
  Galaxies in 3D across the Universe\/} (2015), {\em IAU Symposium\/}, vol.
  309, 190--195.
\newblock \doi{10.1017/S1743921314009648}.

\bibitem[{{Cresci} et~al.(2015{\natexlab{b}}){Cresci}, {Marconi}, {Zibetti},
  {Risaliti}, {Carniani}, {Mannucci} et~al.}]{Cresci:2015ab}
{Cresci} G, {Marconi} A, {Zibetti} S, {Risaliti} G, {Carniani} S, {Mannucci} F,
  et~al.
\newblock {The MAGNUM survey: positive feedback in the nuclear region of NGC
  5643 suggested by MUSE}.
\newblock {\em \aap\/} {\bf 582} (2015{\natexlab{b}}) A63.
\newblock \doi{10.1051/0004-6361/201526581}.

\bibitem[{{Davies} et~al.(2016){Davies}, {Dopita}, {Kewley}, {Groves},
  {Sutherland}, {Hampton} et~al.}]{Davies:2016aa}
{Davies} RL, {Dopita} MA, {Kewley} L, {Groves} B, {Sutherland} R, {Hampton} EJ,
  et~al.
\newblock {The Role of Radiation Pressure in the Narrow Line Regions of Seyfert
  Host Galaxies}.
\newblock {\em \apj\/} {\bf 824} (2016) 50.
\newblock \doi{10.3847/0004-637X/824/1/50}.

\bibitem[{{Bacon} et~al.(2010){Bacon}, {Accardo}, {Adjali}, {Anwand}, {Bauer},
  {Biswas} et~al.}]{Bacon:2010aa}
{Bacon} R, {Accardo} M, {Adjali} L, {Anwand} H, {Bauer} S, {Biswas} I, et~al.
\newblock {The MUSE second-generation VLT instrument}.
\newblock {\em Society of Photo-Optical Instrumentation Engineers (SPIE)
  Conference Series\/} (2010), {\em Society of Photo-Optical Instrumentation
  Engineers (SPIE) Conference Series\/}, vol. 7735, 773508.
\newblock \doi{10.1117/12.856027}.

\bibitem[{{Silk}(2013)}]{Silk:2013aa}
{Silk} J.
\newblock {Unleashing Positive Feedback: Linking the Rates of Star Formation,
  Supermassive Black Hole Accretion, and Outflows in Distant Galaxies}.
\newblock {\em \apj\/} {\bf 772} (2013) 112.
\newblock \doi{10.1088/0004-637X/772/2/112}.

\bibitem[{{Maiolino} and {Rieke}(1995)}]{Maiolino:1995aa}
{Maiolino} R, {Rieke} GH.
\newblock {Low-Luminosity and Obscured Seyfert Nuclei in Nearby Galaxies}.
\newblock {\em \apj\/} {\bf 454} (1995) 95.
\newblock \doi{10.1086/176468}.

\bibitem[{{Risaliti} et~al.(1999){Risaliti}, {Maiolino}, and
  {Salvati}}]{Risaliti:1999aa}
{Risaliti} G, {Maiolino} R, {Salvati} M.
\newblock {The Distribution of Absorbing Column Densities among Seyfert 2
  Galaxies}.
\newblock {\em \apj\/} {\bf 522} (1999) 157--164.
\newblock \doi{10.1086/307623}.

\bibitem[{{Baumgartner} et~al.(2013){Baumgartner}, {Tueller}, {Markwardt},
  {Skinner}, {Barthelmy}, {Mushotzky} et~al.}]{Baumgartner:2013aa}
{Baumgartner} WH, {Tueller} J, {Markwardt} CB, {Skinner} GK, {Barthelmy} S,
  {Mushotzky} RF, et~al.
\newblock {The 70 Month Swift-BAT All-sky Hard X-Ray Survey}.
\newblock {\em \apjs\/} {\bf 207} (2013) 19.
\newblock \doi{10.1088/0067-0049/207/2/19}.

\bibitem[{{Cappellari} and {Copin}(2003)}]{Cappellari:2003aa}
{Cappellari} M, {Copin} Y.
\newblock {Adaptive spatial binning of integral-field spectroscopic data using
  Voronoi tessellations}.
\newblock {\em \mnras\/} {\bf 342} (2003) 345--354.
\newblock \doi{10.1046/j.1365-8711.2003.06541.x}.

\bibitem[{{Vazdekis} et~al.(2010){Vazdekis}, {S{\'a}nchez-Bl{\'a}zquez},
  {Falc{\'o}n-Barroso}, {Cenarro}, {Beasley}, {Cardiel}
  et~al.}]{Vazdekis:2010aa}
{Vazdekis} A, {S{\'a}nchez-Bl{\'a}zquez} P, {Falc{\'o}n-Barroso} J, {Cenarro}
  AJ, {Beasley} MA, {Cardiel} N, et~al.
\newblock {Evolutionary stellar population synthesis with MILES - I. The base
  models and a new line index system}.
\newblock {\em \mnras\/} {\bf 404} (2010) 1639--1671.
\newblock \doi{10.1111/j.1365-2966.2010.16407.x}.

\bibitem[{{Cappellari} and {Emsellem}(2004)}]{Cappellari:2004aa}
{Cappellari} M, {Emsellem} E.
\newblock {Parametric Recovery of Line-of-Sight Velocity Distributions from
  Absorption-Line Spectra of Galaxies via Penalized Likelihood}.
\newblock {\em \pasp\/} {\bf 116} (2004) 138--147.
\newblock \doi{10.1086/381875}.

\bibitem[{{Storchi-Bergmann} and {Bonatto}(1991)}]{Storchi-Bergmann:1991aa}
{Storchi-Bergmann} T, {Bonatto} CJ.
\newblock {Detection of a forbidden O III 5007-A radiation cone in the nuclei
  of NGC 1365 and 7582}.
\newblock {\em \mnras\/} {\bf 250} (1991) 138--143.
\newblock \doi{10.1093/mnras/250.1.138}.

\bibitem[{{Veilleux} et~al.(2003){Veilleux}, {Shopbell}, {Rupke},
  {Bland-Hawthorn}, and {Cecil}}]{Veilleux:2003aa}
{Veilleux} S, {Shopbell} PL, {Rupke} DS, {Bland-Hawthorn} J, {Cecil} G.
\newblock {A Search for Very Extended Ionized Gas in Nearby Starburst and
  Active Galaxies}.
\newblock {\em \aj\/} {\bf 126} (2003) 2185--2208.
\newblock \doi{10.1086/379000}.

\bibitem[{{Sharp} and {Bland-Hawthorn}(2010)}]{Sharp:2010aa}
{Sharp} RG, {Bland-Hawthorn} J.
\newblock {Three-Dimensional Integral Field Observations of 10 Galactic Winds.
  I. Extended Phase (gsim10 Myr) of Mass/Energy Injection Before the Wind
  Blows}.
\newblock {\em \apj\/} {\bf 711} (2010) 818--852.
\newblock \doi{10.1088/0004-637X/711/2/818}.

\bibitem[{{Lindblad}(1999)}]{Lindblad:1999aa}
{Lindblad} PO.
\newblock {NGC 1365}.
\newblock {\em \aapr\/} {\bf 9} (1999) 221--271.
\newblock \doi{10.1007/s001590050018}.

\bibitem[{{Marconi} et~al.(2000){Marconi}, {Oliva}, {van der Werf}, {Maiolino},
  {Schreier}, {Macchetto} et~al.}]{Marconi:2000aa}
{Marconi} A, {Oliva} E, {van der Werf} PP, {Maiolino} R, {Schreier} EJ,
  {Macchetto} F, et~al.
\newblock {The elusive active nucleus of NGC 4945}.
\newblock {\em \aap\/} {\bf 357} (2000) 24--36.

\bibitem[{{Rossa} and {Dettmar}(2003)}]{Rossa:2003aa}
{Rossa} J, {Dettmar} RJ.
\newblock {An H{$\alpha$} survey aiming at the detection of extraplanar diffuse
  ionized gas in halos of edge-on spiral galaxies. II. The H{$\alpha$} survey
  atlas and catalog}.
\newblock {\em \aap\/} {\bf 406} (2003) 505--525.
\newblock \doi{10.1051/0004-6361:20030698}.

\bibitem[{{Harrison} et~al.(2014){Harrison}, {Alexander}, {Mullaney}, and
  {Swinbank}}]{Harrison:2014aa}
{Harrison} CM, {Alexander} DM, {Mullaney} JR, {Swinbank} AM.
\newblock {Kiloparsec-scale outflows are prevalent among luminous AGN: outflows
  and feedback in the context of the overall AGN population}.
\newblock {\em \mnras\/} {\bf 441} (2014) 3306--3347.
\newblock \doi{10.1093/mnras/stu515}.

\bibitem[{{Kewley} et~al.(2001){Kewley}, {Dopita}, {Sutherland}, {Heisler}, and
  {Trevena}}]{Kewley:2001aa}
{Kewley} LJ, {Dopita} MA, {Sutherland} RS, {Heisler} CA, {Trevena} J.
\newblock {Theoretical Modeling of Starburst Galaxies}.
\newblock {\em \apj\/} {\bf 556} (2001) 121--140.
\newblock \doi{10.1086/321545}.

\bibitem[{{Kewley} et~al.(2006){Kewley}, {Groves}, {Kauffmann}, and
  {Heckman}}]{Kewley:2006aa}
{Kewley} LJ, {Groves} B, {Kauffmann} G, {Heckman} T.
\newblock {The host galaxies and classification of active galactic nuclei}.
\newblock {\em \mnras\/} {\bf 372} (2006) 961--976.
\newblock \doi{10.1111/j.1365-2966.2006.10859.x}.

\bibitem[{{Baldwin} et~al.(1981){Baldwin}, {Phillips}, and
  {Terlevich}}]{Baldwin:1981aa}
{Baldwin} JA, {Phillips} MM, {Terlevich} R.
\newblock {Classification parameters for the emission-line spectra of
  extragalactic objects}.
\newblock {\em \pasp\/} {\bf 93} (1981) 5--19.
\newblock \doi{10.1086/130766}.

\bibitem[{{Veilleux} and {Osterbrock}(1987)}]{Veilleux:1987aa}
{Veilleux} S, {Osterbrock} DE.
\newblock {Spectral classification of emission-line galaxies}.
\newblock {\em \apjs\/} {\bf 63} (1987) 295--310.
\newblock \doi{10.1086/191166}.

\end{thebibliography}






\end{document}